PHYSICAL SCIENCES

# Gentle reenergization of electrons in merging galaxy clusters

Francesco de Gasperin,[1,2]* Huib T. Intema,[1] Timothy W. Shimwell,[1] Gianfranco Brunetti,[3] Marcus Brüggen,[2] Torsten A. Enßlin,[4] Reinout J. van Weeren,[1,5] Annalisa Bonafede,[2,3] Huub J. A. Röttgering[1]

Galaxy clusters are the most massive constituents of the large-scale structure of the universe. Although the hot thermal gas that pervades galaxy clusters is relatively well understood through observations with x-ray satellites, our understanding of the nonthermal part of the intracluster medium (ICM) remains incomplete. With Low-Frequency Array (LOFAR) and Giant Metrewave Radio Telescope (GMRT) observations, we have identified a phenomenon that can be unveiled only at extremely low radio frequencies and offers new insights into the nonthermal component. We propose that the interplay between radio-emitting plasma and the perturbed intracluster medium can gently reenergize relativistic particles initially injected by active galactic nuclei. Sources powered through this mechanism can maintain electrons at higher energies than radiative aging would allow. If this mechanism is common for aged plasma, a population of mildly relativistic electrons can be accumulated inside galaxy clusters providing the seed population for merger-induced reacceleration mechanisms on larger scales such as turbulence and shock waves.

## INTRODUCTION

The role of nonthermal components (for example, relativistic particles and magnetic fields) in galaxy clusters is poorly understood because of observational and theoretical difficulties in studying these plasmas on large scales. However, the nonthermal component contributes to the intracluster medium (ICM) pressure biasing hydrostatic mass reconstruction by 10 to 20% (1), and in the era of precision cosmology, this contribution cannot be neglected. A major source of relativistic particles in galaxy clusters is active galactic nuclei that inject cosmic rays (CR) into the diluted thermal gas that pervades the cluster. CR electrons (CRe) emit radio synchrotron emission and hence lose energy. As a result, the CRe become undetectable, and their fate, as well as their possible roles in the cluster evolution, remains uncertain. At the same time, giant radio sources, such as radio relics (2–5) and radio halos (6–8), point to the existence of a widely distributed population of relativistic CRe in the ICM (9, 10).

The currently favored shock models proposed for CRe acceleration in galaxy clusters have found that the acceleration efficiency is very low when electrons are accelerated directly from the thermal pool (11). This low efficiency is hard to reconcile with the observed brightness and radio spectrum of some radio relics that suggest a higher acceleration efficiency (12, 13). Although their direct observation is elusive, a certain amount of CRe with γ ~ 100 must be present in the ICM (14), and a possible solution is that CRe are reaccelerated from a seed population in the ICM (4, 5, 11, 12). Another possibility is that a relatively young cloud of CRe produced by a radio galaxy is reaccelerated through diffusive shock acceleration (15). A few observational pieces of evidence of this last model were recently reported (16–18). However, in the great majority of cases, the presence of a source of CRe near radio relics is missing, leaving unanswered the question of whether clouds of energetic CRe are necessary to power radio relics. A similar problem is present with radio halos that also require an initial reservoir of mildly energetic CRe to reenergize (10).

A related type of radio source is the so-called radio phoenix. Here, the seed CRe filling a lobe of a radio galaxy are only marginally mixed with the ICM. When a shock wave sweeps through the radio bubble, the fossil plasma is compressed adiabatically and reenergized (19). A few observed cases fit the model predictions (20). However, morphological classification of these sources is uncertain, and complementary data such as polarimetric and spectral information are required to reject false positives (21).

Here, we report the first observation of a mechanism that can sustain a population of confined CRe beyond their radiative lifetime. The process of reenergization reported here is so gentle that it barely balances the radiative losses of CR, probing unexplored territories in particle acceleration physics. This mechanism can provide both a long-lived confined cocoon of energetic CRe and, on longer time scales, when the radio plasma mixes with the ICM, a source of seed electrons to power radio relics through DSA and radio halos through turbulence-induced acceleration.

## Abell 1033

Abell 1033 is a moderately massive galaxy cluster ($M_{500} = 3.4 \pm 0.4 \times 10^{14}$ M$_\odot$, $z = 0.1259 \pm 0.0006$) (22) that seems to have experienced a recent merger along an axis oriented north-south and tilted along the line of sight (23). Evidence for a merger is provided by the cluster's thermal x-ray emission. The x-ray emitting gas appears disturbed and shows the presence of a density discontinuity caused by a mild shock wave (Mach = $1.28^{+0.12}_{-0.11}$). Furthermore, the density distribution shows two well-separated peaks oriented north-south, and the redshift distribution is bimodal, both independent indicators of Abell 1033 undergoing a binary cluster merger. The most luminous radio source in the cluster (source S; see Fig. 1) consists of an old pair of lobes from a radio galaxy displaced by the ICM bulk motion. This source is not consistent with a radio relic classification with its observed location, polarization, spectral index, and morphology, and it was classified as a radio phoenix (23).

The presence of radio galaxies with jets strongly bent by their relative motion with respect to the ICM is not uncommon in galaxy clusters and is often associated with cluster mergers (24–26). In Abell 1033, we have two examples of these sources: a head-tail radio galaxy ($z = 0.120$), where the two bent jets merge and become a single flow, and a wide-angle tail (WAT; see Fig. 1) radio galaxy ($z = 0.121$), whose jets are less bent, possibly because the galaxy experiences less ram pressure.

[1]Leiden Observatory, Leiden University, P.O. Box 9513, NL-2300 RA Leiden, Netherlands. [2]Universität Hamburg, Hamburger Sternwarte, Gojenbergsweg 112, D-21029 Hamburg, Germany. [3]Istituto di Radioastronomia, Istituto Nazionale di Astrofisica, Via Piero Gobetti 101, 40129 Bologna, Italy. [4]Max-Planck-Institut für Astrophysik, Karl-Schwarzschild-Str. 1, D-85741 Garching, Germany. [5]Harvard-Smithsonian Center for Astrophysics, 60 Garden Street, Cambridge, MA 02138, USA.
*Corresponding author. Email: fdg@strw.leidenuniv.nl

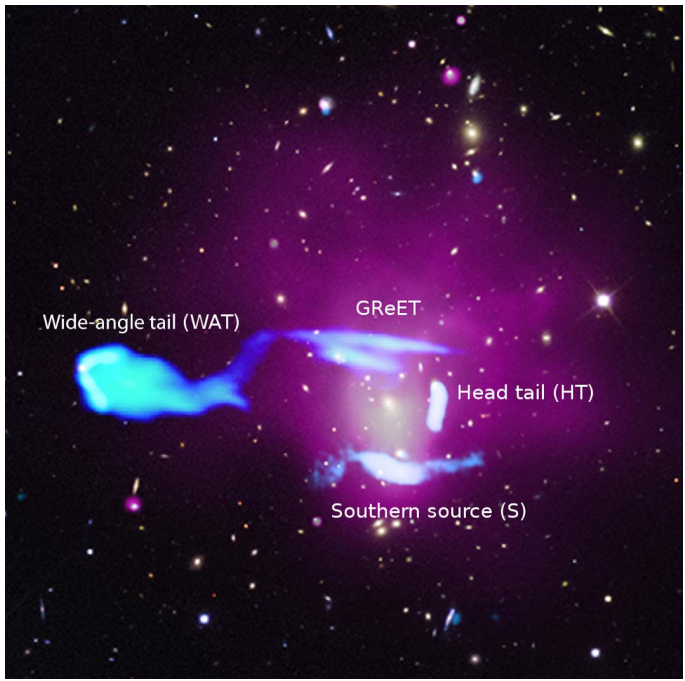

**Fig. 1. Composite image showing optical, radio, and x-ray emission of the galaxy cluster Abell 1033.** The background image shows optical data from the Sloan Digital Sky Survey (*i*, *r*, and *g* filters). In purple, we show the x-ray surface brightness (0.5 to 4 keV, from the Chandra X-ray Observatory) tracing thermal gas, and in blue, we show the radio emission [from Low-Frequency Array (LOFAR)] tracing CR. The source labeled GReET is the main topic of this work. The source labeled Southern source is a probable radio phoenix (23). The image size is around 1 Mpc × 1 Mpc.

**RESULTS**

The focus of this work is the WAT radio galaxy tail, which shows unusual brightness and spectral properties. We have detected the radio tail of this galaxy at 608, 323, 148, and 142 MHz (see Fig. 2). However, only at the lowest frequencies this source appears connected by a bridge to a pair of parallel lines of radio emission that are 320 and 140 kpc long. This peculiar pair of sources, labeled "GReET" (gently reenergized tail) in Fig. 1, is not associated with any optical counterpart and has among the steepest spectrum so far reported for a radio source (spectral index, $\alpha_{323}^{142} = -3.86 \pm 0.03$), defined as $\alpha_{323}^{142} = \log[F_{142\,\text{MHz}}/F_{323\,\text{MHz}}]/\log[142\,\text{MHz}/323\,\text{MHz}]$, where $F_\nu$ is the flux density at frequency $\nu$, which makes it invisible to conventional radio telescopes operating at frequencies higher than a few hundred megahertz. We cannot exclude that the association between the WAT and the GReET through the bridge might be a chance alignment. However, their relation is based not only on their apparent physical connection but also on the same orientation of the WAT and the GReET, the smooth surface brightness, and the smooth variation of spectral index along the tail-bridge-GReET sequence. We believe that an association with the WAT is the most plausible explanation.

The observations suggest that the plasma, ejected through a pair of jets by the active galactic nucleus, was left behind by the galaxy crossing the cluster as two long trails of mildly relativistic electrons ($\gamma \simeq 2000$). The WAT is at a redshift compatible with the northern subcluster, which is closer to the observer than the southern one. When the two clusters started to merge, the WAT continued its motion influenced only by the gravitational potential. On the other hand, the ICM accumulated in the cluster gravitational well, together with the radio-emitting plasma, is slowed down because of ram pressure. If the cluster is in a premerger state and the WAT is part of the northern subcluster, then the displacement of the oldest part of the tails is a natural consequence of the ICM dragging the tails with it. The displacement happens where the ICM density traced by x-ray emission increases (Fig. 1) and reinforces this hypothesis. At the location of the displacement, a bifurcation of the tail seems to be present. It is likely that, in this region, the structure of the tails was modified by the interaction with the ICM. Furthermore, it seems apparent that, although the GReET remained well confined, the more recent emission from the WAT seems to mingle into a single tail. This suggests that the presence of a relevant ICM component is required to keep the plasma confined.

In Fig. 3, we plot the spectral index $\alpha_{323}^{142}$ and the flux density at 142 MHz for a set of regions going from the WAT (the source of the electrons) to the end of the tail. Lines show the predicted spectral index and flux density for a standard plasma aging model [Jaffe-Perola model; (27)] with (i) an injection index equal to −0.65, representing the plasma spectral index when it is injected in the ICM by the active nucleus, therefore before any aging, as measured from high-resolution spectral index map at 610 and 1400 MHz; (ii) a magnetic field $B = 1$, 2.3, or 5 µG; and (iii) negligible adiabatic losses, in line with models where the tail flow is contained and channeled along magnetic field lines (27).

The plot in Fig. 3 shows that the displaced plasma underwent a reenergization process. This process affected both the spectral index and the flux density of the source in the region of the northern trail of the GReET (points 12 to 20), where the extremely steep spectrum of the tail flattens (that is, becomes less negative) at distances >370 kpc. A similar effect is also visible in the southern trail of the GReET. The flattening and brightening imply that the underlying energy distribution of the emitting electrons is boosted toward higher energies. We note that a magnetic field of $B = B_{\text{CMB}}/\sqrt{3} = 2.3$ µG, with $B_{\text{CMB}}$ as the equivalent magnetic field due to inverse Compton (IC) scattering of cosmic microwave background (CMB) photons at the source redshift, maximizes the lifetime of the emitting electrons by minimizing their losses due to synchrotron cooling and IC scattering for an aging plasma observed at a fixed frequency and at Abell 1033's redshift. Higher/lower values of the magnetic field $B$ make the radio source fade faster. At higher field strengths, rapid synchrotron losses quickly deplete CRe. At lower field strengths, we sample CR at a higher Lorentz factor, whose electrons are quickly depleted via synchrotron and IC losses. Therefore, fixing $B = 2.3$ µG gives an upper limit on the average tail radiative age of ~600 My. Assuming negligible projection effects, this implies a galaxy speed of ~730 km/s, which is compatible with the galaxy velocity dispersion of the galaxy cluster (800 ± 80 km/s) (23).

Possible evidence for reenergization of electrons in head tails has been already reported in the literature (28–32). However, our observations at low frequencies allow us to derive the spectrum of the tail at very large distances (long lifetimes), where the spectrum is extremely steep, and highlight the discrepancies between the spectral indices predicted by a minimum aging model and the observed ones with unprecedented evidence. The presence of reenergization is a conclusion that does not depend on the details of our model such as the presence of adiabatic losses or spatial/temporal changes of the magnetic field. We have shown the discrepancy in the context of a scenario that minimizes the energy losses of electrons. Adiabatic losses would make the discrepancy even more severe, whereas sudden variations of the magnetic field experienced by the radiating electrons would flatten the spectrum only locally (that is, for a time scale that is shorter than the electron lifetime), producing a stronger steepening on the longer time scales sampled by our spectral analysis.

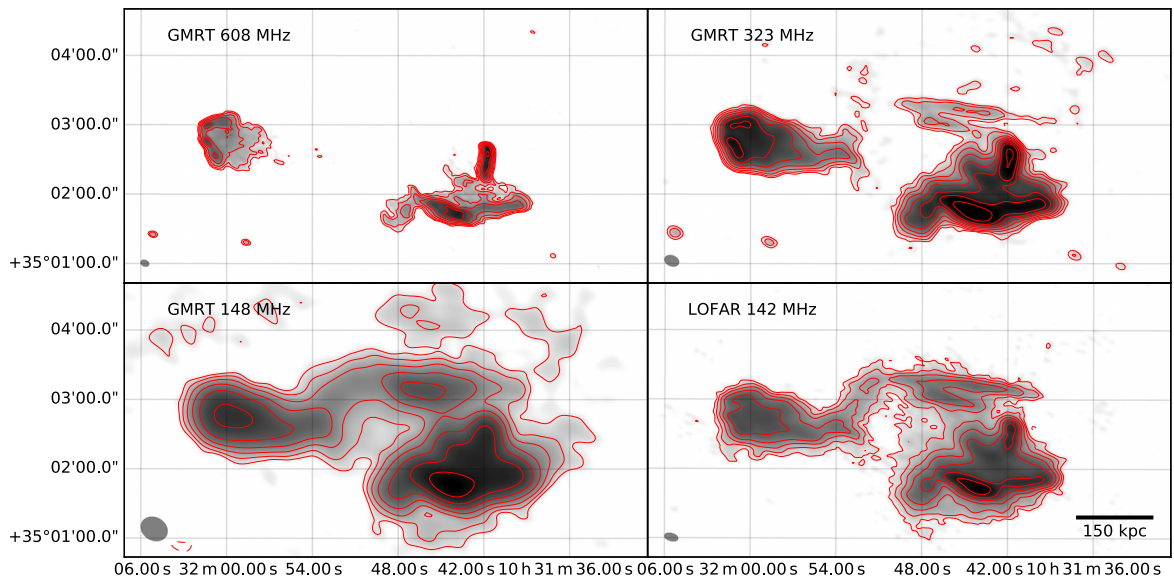

**Fig. 2. Radio images at 608, 323, 148, and 142 MHz. The telescope resolution is shown in the lower left corner of each panel.** Contours are at (1, 2, 4, 8, 16, 32, 64) × 5σ (solid lines) and −5σ (dashed lines), with noises as in Table 1. The GReET is not detected at 608 MHz down to sensitivity of 38 μJy per beam because of its very steep spectrum (α ≃ − 4). The low resolution of the Giant Metrewave Radio Telescope (GMRT) image at 148 MHz does not allow to separate the two trails of the GReET.

At the same time, the observations show something that is completely new: The reenergization must be very gentle. The small spectral flattening implies that the mechanism of reenergization is very inefficient, with a particle acceleration time scale comparable to the radiative loss time scale of the electrons emitting at <100 MHz. Finally, we note that the reenergization must be different from the standard shock reacceleration, which would result in a stronger spectral flattening than observed (18, 33).

### DISCUSSION

Understanding the details of the reenergization mechanism is beyond the aim of this paper. However, given the location of the GReET and the disturbed status of the ICM, we suggest that shocks interacting with the radio tail or turbulence in the tail generated by instabilities driven by the interaction between the tail and the surrounding medium can be responsible for the reenergization of the electrons in the GReET. The challenge is that we observe a spectral index in the GReET that is almost constant for about 300 kpc. This imposes special conditions both in the case of a shock interacting with the tail and in the case that reacceleration is due to turbulence developed within the tail.

If the sound speed in the tail is much greater than that in the surrounding medium, weak shocks may adiabatically compress the aged plasma in the tail, inducing a flattening in the radio spectrum and a brightening of the emission (19, 34, 35). Using the formalism in the study by Enßlin and Gopal-Krishna (19) and starting from an initial spectral index $\alpha_{323}^{142} = -4.5$, the observed flattening of the spectrum would require a very weak shock, with Mach ~ 1.2. A density discontinuity that could be caused by a shock with a similar Mach number was detected in the southern region of the cluster (23). However, particles have different lifetimes along the GReET and thus have different initial spectra when they are shocked. Because the time required by this shock to compress the plasma is much shorter than the lifetime of the GReET, the constant spectral index measured across the source requires very special geometrical and physical configuration. Furthermore, if the shock does not cross the entire GReET at the same time, reenergized particles will restart cooling at different times across the GReET after being compressed by the shock producing a nonuniform spectral index. In principle, a degenerate combination of geometry and variations of the shock Mach number may match with the observations. In the most simple situation, where the shock is moving along the line of sight and crosses the entire GReET at the same time, the Mach number should increase with distance. Specifically, we find that the Mach number should change from 1.2 (at point 13 in Fig. 3) to 1.7 (at point 19). This would require a decline of the upstream temperature in the ICM by a factor of two along the spatial distance covered by the GReET or a change in the velocity of the shock by about 40% across the same spatial scale.

Another possibility is that turbulence developed in the tail due to the interaction with a perturbed surrounding medium can reenergize electrons in the tail. The basic idea is not new, for example, it has been proposed that Rayleigh-Taylor and Kelvin-Helmholtz instabilities in the tails can generate turbulent waves that reaccelerate electrons via second-order Fermi mechanisms (28). However, if true, this possibility would require a substantial revision of the scenario for the evolution of relativistic electrons in these radio sources with respect to what is commonly adopted. In this case, the effect of the reacceleration is to progressively slow down the aging of the electrons until a balance between reacceleration and losses is reached, and a quasi-stationary condition is established around the energy range of electrons that have radiative lifetime comparable to the reacceleration time. As a consequence, a quasi-constant spectrum along the GReET is fairly natural to establish, provided that the properties of the turbulence and properties of the reacceleration mechanism do not change appreciably along the GReET. In this scenario, the challenge is to understand how turbulence and the reacceleration rate are maintained quasi-constant in the tail for a long time scale. The easiest way to circumvent this problem is to assume that turbulence is continuously forced in the tail by the interaction between perturbations in the surrounding medium with the tail itself and that these perturbations are driven in the medium by the cluster dynamics for a time scale and on spatial scales that are larger/comparable to that of the GReET. In Fig. 3, we show the results of a scenario where after 400 My, a reacceleration mechanism becomes operative in the tail (for details, see the Supplementary Materials). The acceleration induces a flattening of the spectrum. Under the assumption that the magnetic field in the tail does not change along its length, the observed synchrotron radiation is determined by the balance between acceleration and energy losses. We find that our data can be explained by assuming very long acceleration times, $\tau_{acc}$ ~700 My; faster accelerations would induce a spectral flattening that is stronger than observed. The modifications of the spectral shape also boost the synchrotron emissivity by a factor of 20 with respect to the pure-aging scenario (Fig. 3). Expectations fall short by a factor of two with respect to the expected boost of brightness. However, projection effects may significantly reduce this discrepancy.

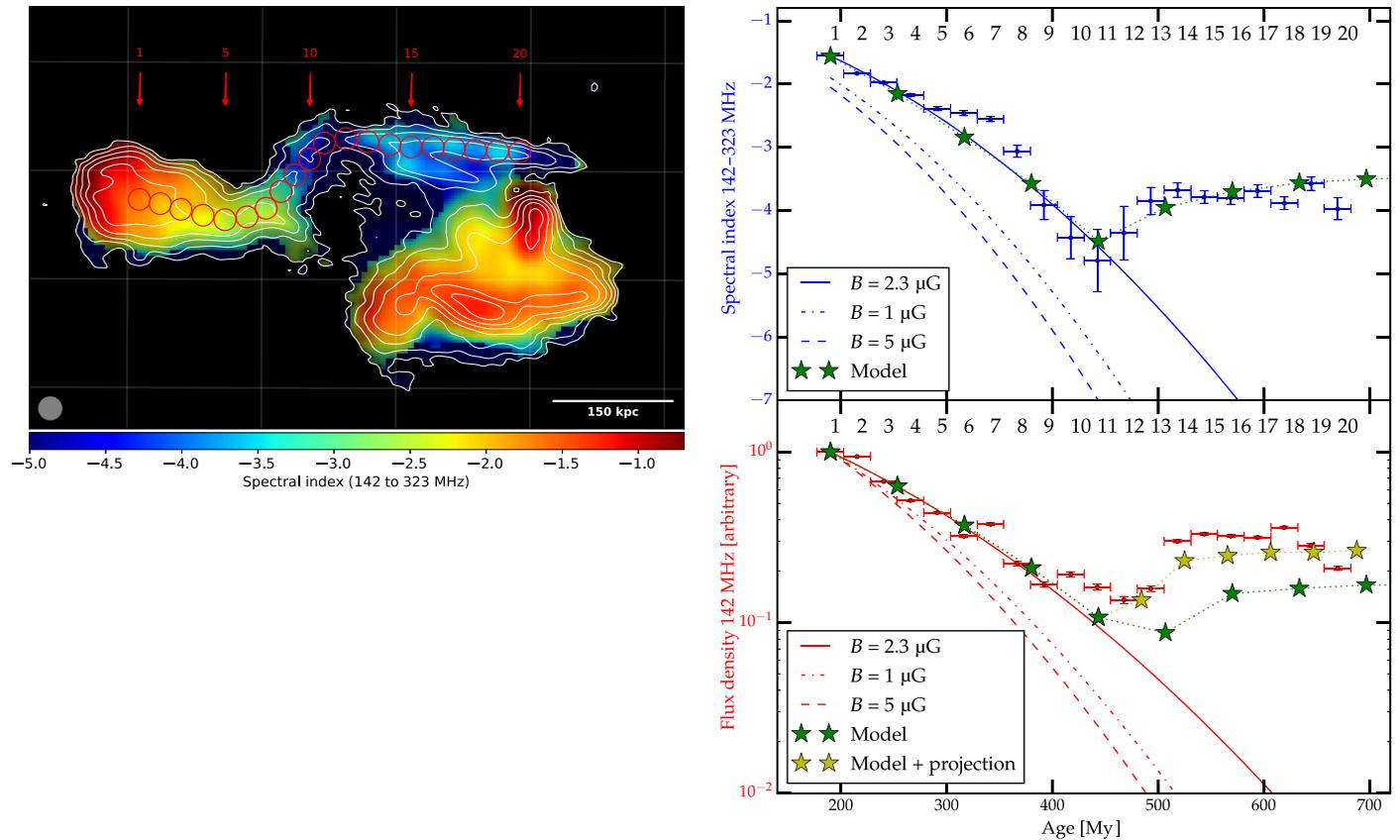

**Fig. 3. Spectral index map of the radio emission in Abell 1033.** (**Left**) Spectral index values (142 to 323 MHz) of the radio emission. Contours are from LOFAR observation at 142 MHz as in the last panel of Fig. 2. To trace the evolution of the spectral index along the tail of the WAT radio galaxy, we defined 20 beam-sized regions. (**Right**) Each point in the plot is associated with one region defined in the left panel. Lines show the spectral index and the 142-MHz flux density predicted by a model, as explained in the text, assuming three possible values for the magnetic field. The position of point 1 has been shifted on the x axis to match the model spectral index for the 2.3-μG case, and the age axis has been stretched to fit the spectral index data (reflecting the unknown galaxy speed). Star markers in the plot show the expected spectral index and flux density if a gentle reacceleration starts at 450 My. Green stars assume no projection effect, while yellow stars assume a 40° inclination of the tail with respect to the line of sight.

We do not include plasma mixing in our models. Mixing alone cannot explain the observed behavior; in particular, it cannot explain the flattening of the spectrum observed between points 10 and 13 in Fig. 3 and the corresponding brightness boost. However, in our confined-plasma model in which particle losses are balanced by gentle reacceleration, spatial mixing may smooth spectral variations and help to maintain a fairly uniform spectral index across the GReET.

The estimate of the acceleration time scale is the most relevant result of our work. It allows us to conclude that the reacceleration must be extremely gentle and that reacceleration mechanisms that are nonconventional (for radio galaxies) must operate within the tail. Understanding the details of the putative reacceleration mechanisms is beyond the aim of this paper. However, one possibility may be betatron reacceleration, more specifically the magnetic pumping mechanism (*36*). In this scenario, magnetic field perturbations are driven continuously into the GReET, creating a time-variable magnetic field that interacts with plasma electrons. In the presence of a source of continuous pitch-angle scattering, electrons will experience a stochastic acceleration via magnetic pumping (quantitative estimates are given in the Supplementary Materials).

Finally, we point out that the long dynamical age estimated for the GReET (600 My), in combination with the width of the trails (25 kpc), allows us to infer the diffusion coefficient due to advection of gas motions around the tail to be $D < l^2/t = (25\ kpc)^2/600\ My = 3.1 \times 10^{29}\ cm^2/s$, implying a velocity $V_t < 120$ km/s of the gas motions on a scale <25 kpc (assuming $D \sim 1/3 V_t \times$ scale). This is comparable with theoretical and observational estimates (*10*, *37*, *38*).

Table 1. Radio observations.

| Telescope | Frequency (MHz) | Observation date | Total observation time (hours) | Bandwidth (MHz) | Resolution | RMS noise (μJy per beam) |
|---|---|---|---|---|---|---|
| LOFAR | 142 | 24 November 2015 | 8 | 48 | 10.3″ × 4.9″ | 180 |
| GMRT | 148 | 26 December 2014 | 8 | 16.4 | 22.2″ × 17.9″ | 800 |
| GMRT | 323 | 2 November 2014 | 5 | 32.8 | 11.2″ × 7.4″ | 42 |
| GMRT | 608 | 2 December 2014 | 4 | 32.8 | 5.8″ × 3.7″ | 38 |

### CONCLUSIONS

Here, we reported the observational evidence of a mechanism that can reenergize electrons in radio galaxy tails. We named the revived source GReET. This mechanism seems to be activated when the radio tail crosses the disturbed region of the merging cluster Abell 1033. Two scenarios were investigated: reenergization due to compression by very weak shocks or reenergization due to stochastic reacceleration mechanisms driven into the tail. Although both mechanisms can explain the observational data, the compression scenario requires a precise geometrical tuning.

Regardless of the energization mechanism, if this gentle reenergizing process observed in Abell 1033 is common in tails of radio galaxies in galaxy clusters, then it has strong consequences on the life cycle of relativistic particles in radio galaxies and of fossil radio plasma in the ICM as well as on the formation scenarios of cluster-scale radio emission. If electrons released by radio galaxies in the ICM could live as long as seen in the case of Abell 1033 (>0.5 billion years), then they would be able to accumulate in larger quantities and with higher energies than previously thought. This could produce a seed population of energetic particles for merger-induced reacceleration mechanisms, such as turbulence and shocks, that were proposed to explain cluster-scale radio sources. At the same time, the scenario predicts the presence of a population of very steep spectrum tails in dynamically disturbed clusters that would light up at low radio frequencies, as found in recent LOFAR observations (*39*).

### MATERIALS AND METHODS
#### GMRT data reduction

For each observing frequency, the visibility data were processed using an Astronomical Image Processing System (AIPS)–based (AIPS: www.aips.nrao.edu) pipeline incorporating SPAM [Source Peeling and Atmospheric Modeling (*40*)] ionospheric calibration. The AIPS-based, semiautomated pipeline processing is very similar for each of the GMRT observing frequencies, being different only in the detailed settings of automated flagging. The pipeline uses the ParselTongue interface (*41*) to access AIPS tasks, files, and tables from Python. Flux and bandpass calibrations were derived from 3C 147 after three iterations of flagging and calibrating based on the model by Scaife and Heald (*42*). Additionally, instrumental phase calibrations were determined by filtering out ionospheric contributions (*40*), an important step for direction-dependent ionospheric calibration later on. Calibrations were transferred and applied to the target field data, simple clipping of spurious visibility amplitudes was applied, and data were averaged in time and frequency and converted to Stokes *I* to speed up processing. The effective bandwidths after flagging and averaging are 16.4, 32.8, and 32.8 MHz, centered on 148, 323, and 608 MHz, respectively. Self-calibration of the target field was started with an initial phase calibration using a multi point source model derived from the NRAO VLA (National Radio Astronomy Observatory Very Large Array) Sky Survey (NVSS), Westerbork Northern Sky Survey (WENSS), and Very Large Array Low-Frequency Sky Survey (VLSS) radio source catalogs (*43*–*45*), followed by (facet-based) wide-field imaging and CLEAN deconvolution of the primary beam area and several bright outlier sources (out to five primary beam radii). The visibility weighting scheme used during imaging is a mixture between uniform and robust weighting (AIPS ROBUST -1), which generally provides a well-behaved point spread function (without broad wings) by down-weighting the very dense central *uv*-coverage of the GMRT. Self-calibration was repeated three more times, including amplitude calibration in the final round and outlier flagging on residual (source model subtracted) visibilities in between imaging and calibration.

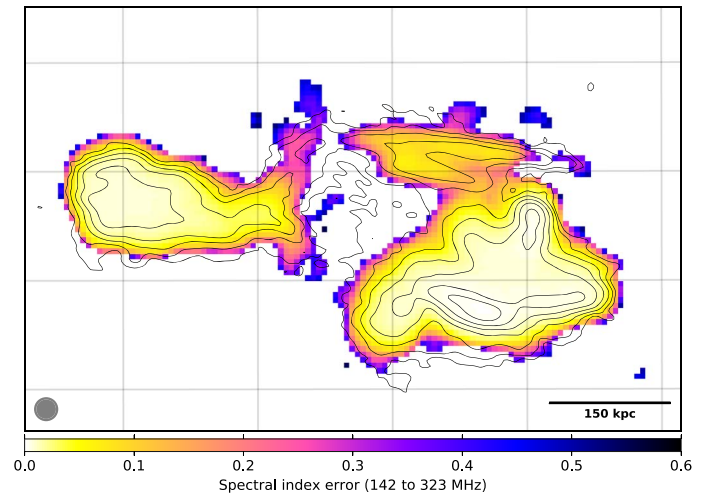

**Fig. 4. Spectral index error map.** Each pixel shows the standard deviation of the bootstrapped spectral index distribution.

Low-frequency radio observations are affected by ionospheric-induced delay caused by a nonuniform refractive index above the array. This effect is inversely proportional to the observing frequency and strongly direction-dependent. As a consequence, calibration strategies for low-frequency observations need to correct for it. For our GMRT data, this correction was carried out with two rounds of SPAM calibration (*40*, *46*). In each round, (i) bright sources in and around the primary beam area were peeled [for example, see the study of Noordam (47)], (ii) the peeling phase solutions were fitted with a time-variable two-layer phase screen model, (iii) the model was used to generate ionospheric phase correction tables for the grid of viewing directions defined by the facet centers, and (iv) the target field was reimaged and deconvolved while applying the appropriate correction table per facet. Resolutions and noise levels are listed in Table 1.

### LOFAR data reduction

Abell 1033 was observed with the LOFAR High-Band Antenna, in "Dual inner" mode, on 24 November 2015 (observation ID: 413284) as part of the ongoing LOFAR Two-meter Sky Survey [LoTSS; (48)]. The observed band was continuous from 120 to 168 MHz. The calibrator source 3C 196 was observed for 10 min before the observation of the target field. The standard LOFAR direction-independent calibration pipeline (https://github.com/lofar-astron/prefactor) was used to measure systematic effects on the calibrator, transfer the calibration solutions to the target, correct for the average ionospheric conditions in the target field, and subtract all the sources from the field to leave an empty $uv$–data set. The recently developed LOFAR direction-dependent calibration pipeline (https://github.com/lofar-astron/factor) (49) was used to add back the sources within approximately 0.5° of Abell 1033 and to self-calibrate these data to correct for the ionospheric and beam errors in this direction. Primary beam–corrected images of Abell 1033 were produced at a resolution of 10.3″ × 4.9″ with a sensitivity of 180 μJy per beam.

### Spectral index

The spectral index map was generated by combining the LOFAR image at 142 MHz and the GMRT image at 323 MHz and by cutting the minimum $uv$-range to 100λ in both cases. The two images were convolved to the maximum common resolution of 12″ and regridded to the same pixelation. For each pixel where both images have a signal >3σ, we calculated the local spectral index value. The linear regression was made with a bootstrap least-squares algorithm that takes into account flux density errors (local RMS noise). The spectral index of each pixel was estimated 1000 times, and the mean of the resulting sample was used as an estimator of the spectral index. In Fig. 4, we show the standard deviations of the distributions. For extended regions, the spectral index was calculated in the same manner, but the flux density errors were computed as $S_{err} = \sigma \cdot \sqrt{N_{beam}}$, where σ is the local image RMS noise, and $N_{beam}$ is the number of beams covering the extension of the region. To further test the reliability of our spectral index estimations, we recalculated the values used in the top right plot in Fig. 3 using GMRT maps at 148 and 323 MHz. We obtained compatible results.

### SUPPLEMENTARY MATERIALS

Supplementary material for this article is available at http://advances.sciencemag.org/cgi/content/full/3/10/e1701634/DC1
Calculations
Reference (50)

**Acknowledgments:** We would like to thank the staff of the GMRT who made these observations possible. GMRT is run by the National Centre for Radio Astrophysics of the Tata Institute of Fundamental Research. This paper is based, in part, on data obtained with the International LOFAR Telescope (ILT). LOFAR is designed and constructed by ASTRON. It has facilities in several countries that are owned by various parties (each with their own funding sources) and are collectively operated by the ILT foundation under a joint scientific policy. **Funding:** F.d.G. is supported by the VENI research program with project number 1808, which is financed by the Netherlands Organisation for Scientific Research. R.J.v.W. is supported by a Clay Fellowship awarded by the Harvard-Smithsonian Center for Astrophysics. G.B. acknowledges partial support from grant PRIN Istituto Nazionale di Astrofisica 2014. A.B. acknowledges support from the ERC-Stg 714245 DRANOEL. **Author contributions:** F.d.G.: Principal Investigator (PI) GMRT observations, data analysis, and paper writing. H.T.I.: GMRT data reduction and data interpretation. T.W.S.: LOFAR data reduction. G.B.: theoretical modeling. M.B.: data interpretation. T.A.E.: theoretical modeling. R.J.v.W.: LOFAR data reduction. A.B.: data interpretation. H.J.A.R.: PI LOFAR observations. F.d.G.: claims responsibility for all images. **Competing interests:** The authors declare that they have no competing interests. **Data and materials availability:** All data needed to evaluate the conclusions in the paper are present in the paper and/or the Supplementary Materials. Additional data related to this paper may be requested from the authors.

Submitted 16 May 2017
Accepted 13 September 2017
Published 4 October 2017
10.1126/sciadv.1701634

**Citation:** F. de Gasperin, H. T. Intema, T. W. Shimwell, G. Brunetti, M. Brüggen, T. A. Enßlin, R. J. van Weeren, A. Bonafede, H. J. A. Röttgering, Gentle reenergization of electrons in merging galaxy clusters. *Sci. Adv.* **3**, e1701634 (2017).